\documentclass[pra,twocolumn,notitlepage]{revtex4-1}

\usepackage{amsthm,amssymb,mathrsfs,setspace,times}

\usepackage{amsmath}
\usepackage{graphicx}

\begin{document}

\title{Non-Markovian open dynamics from collision models}

\author{Vijay Pathak}
\affiliation{School of Physics, IISER Thiruvananthapuram, Kerala, India 695551}
\email{vijayp@iisertvm.ac.in}

\author{Anil Shaji}
\affiliation{School of Physics, IISER Thiruvananthapuram, Kerala, India 695551} 
\email{shaji@iisertvm.ac.in}

\begin{abstract}
	Convex combinations of the completely positive (CP) as well as CP-divisible, continuous time dynamical maps arising from collision models are investigated. While the individual maps are both CP and Markovian we find that convex combinations remain CP but not necessarily Markovian. Examples of such combinations for qubit dynamical maps arising from collisional models are worked out and the invertibility, CP-divisibility, P-divisibility as well as Markovian properties of such maps are explored. 
\end{abstract}

\maketitle

The seminal paper by Gorini, Kossakowski and Sudarshan~\cite{Gorini:1975dn} established the general form of the generator of a completely positive dynamical semi-group describing the Markovian dynamics of an open quantum system. The structure and form of the master equation pointed out the (Lindblad) operators\footnote{These operators are referred to as {\em Lindblad operators} owing to a paper on the same subject submitted almost concurrently by G.~Lindblad~\cite{Lindblad:1976de, Chruscinski:2017fp}.} and their coefficients that need to be identified and measured experimentally in order to obtain a complete description of the open dynamics of a quantum system. Such reconstruction of the dynamics goes by the name of quantum process tomography in recent literature and it is rapidly becoming a mature field involving sophisticated experimental and theoretical methods~\cite{Nielsen00a,chuang97a,Howard06,PhysRevA.67.042322, ziman06,Gaikwad:2018gh,Pollock:2018cd}. A constraint and a challenge both have to be addressed when writing down a master equation in the Gorini-Kossakowski-Sudarshan-Lindblad (GKSL) form to describe the open dynamics of a quantum system of interest. The constraint is that the master equation can describe only Markovian open dynamics of the system. The challenge is that the operators and the coefficients appearing in the equation have to satisfy the requirement of complete positivity. There is quite a bit of discussion in the literature on whether complete positivity is required for open quantum evolution~\cite{Buscemi:2014dc, jordan04a, jordan04b, pechukas94a, shaji05a, Joseph:2018kl, Vacchini:2016vf}. Similarly there quite a bit of interest in constructing useful master equations that describe non-Markovian open dynamics as well~\cite{breuer02a, breuer1999stochastic, breuer2009structure, deVega:2017gu, kretschmer2016collision, lorenzo2017quantum, PhysRevLett.112.120404, RevModPhys.88.021002, rivas2012open, suarez92a}.

Open dynamics of a quantum system of interest involves its interaction with a well defined environment whose properties or dynamics are neither known nor accessible to the observer. Let us denote the system as $S$ with its state at time $t$ being described by the density matrix $\rho_S(t)$. The state of the environment $E$ is $\rho_E(t)$ and the joint state of the two is denoted as $\rho_{SE}(t)$. Starting from an initial state $\rho_S(0)$, the state of $S$ at a later time is given by 
\begin{equation}
	\label{SEevolve}
	\rho_S(t) = {\rm Tr}_E [ U_{S\!E}^{\phantom \dagger}(t) \rho_{S\!E}(0) U_{S\!E}^\dagger(t) ],
\end{equation}  
where $U_{S\!E}(t)$ is the unitary time evolution operator acting on the joint system $S\!E$ and $\rho_S(0) ={\rm Tr}_E [\rho_{S\!E}(0)]$. Implicit in this statement is the assumption that $E$ is chosen such that $S\!E$ constitutes an isolated system. The finite time evolution of the system from $\rho_S(0)$ to $\rho_S(t)$ can be described by a dynamical map~\cite{PhysRev.121.920,choi72a,choi75,stormer63} such that
\begin{equation}
	\label{map}
	\rho_S(t) = \Lambda_t \rho_S(0).
\end{equation}
In the following, we are interested in cases where the dynamical map has the semigroup property, 
\begin{equation}
	\label{eq:semigroup}	
	\Lambda_s \cdot \Lambda_t = \Lambda_{t+s}.
\end{equation}
The semigroup property allows one to write a differential equation describing the evolution of the open system in continuous time as
\begin{equation}
	\label{semigroup}
		\frac{d \rho_S(t)}{dt} = {\mathcal L} \rho_S(t),
\end{equation}
where ${\mathcal L}$ is the generator of the dynamical semigroup. The GKSL master equation is a special case of such a differential equation valid under the additional assumption of complete positivity. 

The construction in Eq.~(\ref{SEevolve}) leads to a completely positive (CP) dynamical map defined on the set of all states of $S$ provided $\rho_{S\!E}(0) = \rho_S(0) \otimes \rho_E$. The GKSL equation demands complete positivity at every instant which means that $\rho_{S\!E}(t) = \rho_S(t) \otimes \rho_E(t)$ for all values of $t$. When viewed in conjunction with the fact that the subset of states of a bipartite system - in this case $S$ and $E$ - that do not share any non-classical correlations is a set of measure zero within the set of all states of $S\!E$~\cite{Vedral}, this condition may appear quite stringent but the requirement of complete positivity is one of the key ingredients that allows one to identify the general form of the generator of the dynamical semigroup~\cite{Chruscinski:2017fp}. The semigroup condition is equivalent to the Markov property that demands that the environment be memoryless. In other words $E$ does not change its state in a manner that depends on the evolution of $S$. In practice the Markov assumption is brought in by essentially associating a stationary state with $E$ so that $\rho_{S\!E}(t) = \rho_S(t) \otimes \rho_E$. This assumption can be justified when $E$ is much larger compared to $S$ or if the relaxation time scale for $E$ is much shorter compared to the typical time scale of evolution of $S$. A weak coupling between $S$ and $E$ also can be shown to produce Markovian open dynamics for $S$~\cite{breuer02a}. 

When the environment is large and complex, approximate stationarity of its state relative to the dynamics of the system is a reasonable assumption. However with increasing access and control over individual atoms, photons, electrons and other quantum systems, we find that the effective environment with which they interact over the time scales of interest is in itself typically quantum or at best mesoscopic in nature and size. For concreteness one may imagine an atom in a cavity interacting with a few cavity modes serving as its environment or an ion in a trapped array interacting with a few phonon modes. The open evolution of $S$ becomes manifestly non-Markovian in these scenarios and consequently there is quite a bit of interest in constructing useful master equations describing such evolution~\cite{RevModPhys.88.021002}.  A general non-Markovian master equation has the form 
\begin{equation}
	\frac{d \rho_S(t)}{dt} = \int_0^t K(t-v) \rho_S(v) \, dv.
\end{equation}
Only very specific choices of the memory kernel $K(t-v)$ however leads to completely positive evolution in the non-Markovian case. Often even on memory kernels constructed based on very plausible physical assumptions, additional constraints have to be imposed to guarantee complete positivity~\cite{breuer2009structure, PhysRevLett.101.140402}. Finding the most general form of the memory kernel that leads to completely positive evolution is a problem that still remains open. 

Collision models~\cite{Sudarshan:2003hm, Vacchini14a, ziman2005description, Ciccarello:2013ea, ciccarello2013collision, Filippov:2017fr, kretschmer2016collision, lorenzo2017quantum, Rybar:2012bn}  provide a means of constructing master equations that are guaranteed to be completely positive. In collision models, the environment is modeled as a stream of ancillary quantum systems each of which interact briefly, sequentially and independently with the open system. The ancilla are typically chosen to be low dimensional quantum systems like qubits. The interaction is unitary in nature and the interaction time is assumed to be brief.  Prior to each `collision' with the environment the joint state of the system and the ancilla with which it is going to interact is a simple product ensuring that the brief coupled unitary evolution of the two leads to completely positive reduced dynamics for $S$. A master equation is obtained formally in the limit in which the number of collisions goes to infinity while simultaneously the duration of each collision goes to zero. Alternatively one can consider the continuous time dynamics that interpolates smoothly between the discrete state changes produced by each collision with the environment. The master equation obtained using the collision model approach is not just CP but by construction it is CP-divisible as well which implies that all the three maps appearing in Eq.~(\ref{eq:semigroup}) are completely positive.

We consider here convex combinations of dynamical maps, each of which is generated by a suitable collision model. We restrict to maps on single qubits. We observe that the resultant maps are completely positive due to the manner in which it is constructed. When the dynamical maps are invertible, we can also find corresponding generators for the map from which a master equation can be written down. We show that in general the master equation is not CP divisible or sometimes not even P-divisible and it describes non-Markovian evolution even if the individual collision models that went into the construction of the master equation are themselves CP divisible and Markovian. The main advantage of constructing a non-Markovian master equation in this manner is that it is guaranteed to lead to completely positive evolution. The construction can conversely be viewed as providing an unravelling of a non-Markovian, completely positive master equation in terms of a collection of Markovian, CP-divisible ones. 

This paper is structured as follows. In the next section we briefly recap the collision model construction and discuss our general results. In the following section we present an illustrative example in detail. The last section contains a brief discussion of our results.  

\section{Convex sums of collision model based dynamical maps \label{sec2}}

The collision model introduced in~\cite{ziman2005description} gives a microscopic picture for a Markovian master equation. The environment of the open system of interest is modeled as a collection of $n$ identical `particles' that `collide' sequentially with the system. The initial state of each particle is denoted as $\xi_{\vec{u}}$ where $\vec{u}$ is a collection of parameters that characterize the initial density matrix $\xi$ of each environment particle. The overall initial state of the environment is thus $\xi_{\vec{u}}^{\otimes n}$. Each collision is described by a unitary transformation $U_{\vec{\eta}}$ acting on $\rho_S^{(i)} \otimes \xi_{\vec{u}}$ where $\rho_S^{(i)}$ is the state of the system after the $i^{\rm th}$ collision. The parameters $\vec{\eta}$ define the unitary transformation in terms of a fixed operator basis. So we have
\[ \rho_S^{(i+1)} = {\rm Tr}_{E} \big[ U_{\vec{\eta}}^{\vphantom{\dagger}} \,\rho_S^{(i)} \otimes \xi_{\vec{u}} \, U_{\vec{\eta}}^{\dagger} \big], \]
where the trace is over the state of the $(i+1)^{\rm th}$ environment particle. Since all environment particles are identical, each collision induces a dynamical map on the state space of the system denoted by ${\mathcal E}(\vec{u},\vec{\eta})$ such that
\begin{equation}
	\label{collisionmap1}
	\rho_S^{(i+1)} = {\mathcal E}\rho_S^{(i)},
\end{equation}
where we have chosen not to show explicitly the dependence of ${\mathcal E}$ on the parameters $\vec{\eta}$ and $\vec{u}$ of the unitary and the state of the environment particles respectively. The map ${\mathcal E}$ is CP by construction and the map corresponding to $n$ collisions with the environment particles is easily seen to be $({\mathcal E})^n \equiv {\mathcal E}_n$. The dynamical semigroup property is automatically satisfied in that
\[ {\mathcal E}_m {\mathcal E}_n = {\mathcal E}_{n+m}. \]
We can define a family of dynamical maps indexed by a continuous parameter $t$ by interpolating between the sequence of discrete transitions described by ${\mathcal E}_n$ by identifying $t = n\tau$ where $\tau$ represents the `typical' time for each collision. Formally, for the interpolation, one takes the limit of $\tau$ going to zero with the number of collisions $n$ going to infinity while keeping $t$ finite. The continuous time dynamical map ${\mathcal E}_t$ coincides with ${\mathcal E}_n$ at discrete points $n = t/\tau$ and ${\mathcal E}_0 \equiv {\mathbb I}$ (the identity map). 

Provided ${\mathcal E}_t^{-1}$ exists, we can compute the generator corresponding to the map as~\cite{ziman2005description},
\begin{equation}
	\label{eq:generator}
	{\mathcal L}_t = \frac{d{\mathcal E}_t}{dt} {\mathcal E}_t^{-1}, 
\end{equation}
leading to a dynamical equation for the state of $S$ with the same form as in Eq.~(\ref{semigroup}). The divisibility property of the dynamics allows for expression of the time evolution of the system in the GKSL form, 
\begin{eqnarray}
	\label{GKSL1}
	\frac{d\rho_S(t)}{dt} & = &  -i\big[H_t,\, \rho_S(t)\big] \nonumber \\
	&& \quad + \; \frac{1}{2} \sum_{\alpha, \beta} c_{\alpha,\beta} (t)\Big( \big[ \Lambda_\alpha^t, \, \rho_S(t) \Lambda_\beta^t \big] \nonumber \\
	&&  \qquad \qquad \qquad + \; [\Lambda_\alpha^t \rho_S(t), \, \Lambda_\beta^t \big] \Big),
\end{eqnarray}
where $\Lambda_\alpha^t$ furnish a suitable trace-orthogonal operator basis with 
\[ H_t = \sum_\alpha h_\alpha (t)\Lambda_\alpha^t.\]
 If the generator ${\mathcal L}_t$ is not time dependent, then the coefficients $c_{\alpha, \beta}(t)$ and $h_\alpha(t)$ are also time independent and Eq.~(\ref{GKSL1}) reduces to the standard Markovian master equation obtained in~\cite{Gorini:1975dn,Lindblad:1976de}.
 
 The continuous time dynamical map ${\mathcal E}_t$ obtained from the collision model depends on both the state of the environment particle determined by the parameters $\vec{u}$ as well as the the nature of the interaction between the system and each environment particle specified by $\vec{\eta}$. We consider an ensemble of possible states for the environment particles distributed according to the probability distribution $p(\vec{u})$. We keep the interaction between the system and the environment particles the same irrespective of the state of the environment particle. We look in particular at continuous time dynamical maps of the form
\begin{equation}
	\label{eq:SumMap1}
	\widetilde{\mathcal E}_t(\vec{\eta}) = \int d\vec{u}\, p(\vec{u}) {\mathcal E}_t (\vec{u}, \vec{\eta}). 
\end{equation}
In what follows will avoid writing explicitly the dependence of the maps on $\vec{\eta}$ since we are assuming it to be same for collisions with all types of environment particles. 

Since ${\mathcal E}_t$ are CP dynamical maps, they allow an operator sum (Kraus)~\cite{PhysRev.121.920, kraus71a, kraus83} representation
\[ {\mathcal E}_t[\rho] = \sum_{j=1}^{n} A_j \rho A_j^\dagger, \qquad \sum_j A_j^\dagger A_j = {\mathbb I}. \]
Substituting this in Eq.~(\ref{eq:SumMap1}), we obtain
\[ \widetilde{\mathcal E}_t[\rho] = \int d\vec{u} \, p(\vec{u}) \, \sum_{j=1}^{n(\vec{u})}  A_j (\vec{u}) \rho A_j^\dagger (\vec{u}).\]
The number of terms in the operator sum representation is not the same for all maps. We can however eliminate the $\vec{u}$-dependence of the upper limit of the sum in the equation above by choosing the limit to be equal the $n(\vec{u})$ for the particular ${\mathcal E}_t(\vec{u})$ with the maximum number of Kraus operators. Let this number be denoted by $n_{\rm max}$. The set Kraus operators $\{A_j(\vec{u})\}$ for all other maps can be padded with sufficient number of null operators $0_j$ so that all the maps have the same number of terms in their respective operator sum representations. So we obtain
\begin{eqnarray}
	\label{compositeCP}
	 \widetilde{\mathcal E}_t[\rho] & = & \sum_{j=1}^{n_{\rm max}}  \int d\vec{u} \, p(\vec{u}) \,  A_j (\vec{u}) \rho A_j^\dagger (\vec{u})\nonumber \\
	 &  = & \sum_{j=1}^{n_{\rm max}}  \int d\vec{u}   \,  A_j' (\vec{u}) \rho A_j'^{\dagger} (\vec{u}),
\end{eqnarray}
where $A_j' (\vec{u}) \equiv \sqrt{p(\vec{u})} A_j (\vec{u})$. We also have
\begin{eqnarray*}
	\sum_{j=1}^{n_{\rm max}}  \int d\vec{u} \,   A_j'^{\dagger} (\vec{u})  A_j' (\vec{u}) & = &  \int d\vec{u} \, p(\vec{u}) \, \sum_{j=1}^{n(\vec{u})} A_j^\dagger(\vec{u}) A_j(\vec{u}) \\
	& = &   \int d\vec{u} \, p(\vec{u}) \, {\mathbb I} = {\mathbb I}.
\end{eqnarray*} 
Treating the integral in Eq.~(\ref{compositeCP}) as a sum of infinitesimal increments, we see that $ \widetilde{\mathcal E}_t$ also has an operator sum representation showing that this map is also CP given that all ${\mathcal E}_t(\vec{u})$ are also CP.  However the convex combinations of CP maps need not necessarily describe Markovian open evolution~\cite{Filippov:2017fr}. In the following we examine the properties of such dynamical maps acting on qubit states.

\subsection{Qubit maps \label{qubitmaps}}

We now restrict our discussion to dynamical maps on qubits arising from collision models. The state of $S$ is written as $\rho_S = ({\mathbb I} + \vec{r} \cdot \vec{\sigma})/2$ in terms of the Pauli matrices $\sigma_j$, $j=1,2,3$ that, along with the identity operator, ${\mathbb I}$, furnish an operator basis for $SU(2)$ matrices. Any dynamical map can be expressed as a transformation on the  vector $(1, \vec{r})$ as
\[ {\mathcal E}_t = \left( \begin{array}{cccc}
	1 & 0 & 0 & 0 \\
	s_1 & T_{11} & T_{12} & T_{13} \\
	s_2 & T_{21} & T_{22} & T_{23} \\
	s_3 & T_{31} & T_{32} & T_{33} 
\end{array} \right) \equiv 
\left( \begin{array}{cc}
	1 & \vec{0} \\
	\vec{s} & T 
\end{array} \right) \]
The generic dynamical map implements an affine transformation on the Bloch vector $\vec{r}$ of the qubit state as 
\[ \vec{r} \rightarrow \vec{r}' = T\vec{r} + \vec{s}. \]
The generator of the transformation has the form~\cite{ziman2005description},
\[ {\mathcal L}_t = \frac{d{\mathcal E}_t}{dt}  {\mathcal E}_t^{-1} =  \left( \begin{array}{cc}
	0 & \vec{0} \\
	\vec{l} & L 
\end{array} \right). \]
The corresponding master equation can be written in the standard operator basis of the Pauli matrices as
\begin{eqnarray}
	\label{eq:GKSL2}
	\frac{d \rho_S(t)}{dt} & = &  -i \big[ H_t, \, \rho_S(t) \big] \nonumber \\
	&& \quad + \; \frac{1}{2} \sum_{j,k=1}^3 c_{jk}(t) \Big( \big[ \sigma_j, \, \rho_S(t) \sigma_k \big] \nonumber \\ && \qquad \qquad \qquad + \; \big[\sigma_j \rho_S(t), \, \sigma_k \big] \Big),
\end{eqnarray}
with $H = \sum h_j(t) \sigma_j$. The matrix elements of the generator and the coefficients appearing in Eq.~(\ref{eq:GKSL2}) are connected by the relations,
\begin{eqnarray}
	\label{eq:GKSL3}
	h_j & = & \frac{1}{4} \epsilon_{jkl} ({\mathcal L}_{lk} - {\mathcal L}_{kl} ), \nonumber \\
	c_{jj} & = & \frac{1}{4}  ({\mathcal L}_{jj} - {\mathcal L}_{kk} - {\mathcal L}_{ll}), \nonumber \\
	c_{jk} & = & \frac{1}{4} ({\mathcal L}_{jk} + {\mathcal L}_{kj} - i \epsilon_{jkl} {\mathcal L}_{l0}),
\end{eqnarray}
where $\epsilon_{jkl}$ is the fully antisymmetric tensor with $c_{kj} = c_{jk}^*$.

We further assume that the environmental particles are qubits. Then $\vec{u}$ becomes the Bloch vector specifying the state of these particles. For every choice of $\vec{u}$ we have a continuous time map ${\mathcal E}_t(u_x, u_y, u_z, \vec{\eta})$ obtained by considering a sequence of collisions of $S$ with the environment qubits. Following~\cite{ziman2005description}, we take the interaction between the system and each of the environment qubits of the form
\begin{equation}
	\label{eq:unitary}
	U_\eta = \cos \eta \, {\mathbb I} + i \sin \eta W,
\end{equation}  
where $W$ is a fixed two qubit unitary. We therefore have $\vec{\eta} = \eta$. The collision leaves $S$ unchanged with probability $\cos^2 \eta$ and with probability $\sin^2 \eta$ a fixed two qubit unitary $W$ is applied. As described earlier we consider maps of the form
\[ \widetilde{\mathcal E}_t (\eta) = \int d^3 u \, p(u_x, u_y, u_z) \, {\mathcal E}_t(u_x, u_y, u_z, \eta),\]
with the aim of obtaining the generators and corresponding master equations for such convex sums of CP-divisible dynamical maps. Here $p(u_x, u_y, u_z)$ may be visualized a probability distribution over the Bloch sphere of the environment qubit.  We explore the nature of the master equations obtained corresponding to different choices of $p(u_x, u_y, u_z)$ and we find that in general the open evolution of $S$ described by these master equations is non-Markovian.

\section{Example: Homogenization \label{sec3}}

Homogenization describes a process by which every state, $\rho_S$, of the system is transformed into a fixed state $\xi$ which also is the state of each of the environment qubits. This process was considered in detail as a collision model in~\cite{ziman2005description}. Homogenization is possible exactly only in the asymptotic limit where the number of collisions with the environment, $n$, goes to infinity. For finite $n$ the process is still invertible and maps the single qubit Bloch sphere into an arbitrarily small ball around $\xi$. Homogenization is implemented through the partial-swap operation:
\begin{equation}
\label{eq:swap}
U_{\eta} = \cos \eta \ I +  i \sin \eta \ S,
\end{equation}
where $S$ is the swap operator, 
\[ S\vert a\rangle_S \otimes\vert b\rangle_E = \vert b\rangle_S \otimes\vert a\rangle_E, \]
for all $\vert a\rangle_S$ and $\vert b\rangle_E$. The unitary swaps the system-ancilla state with the probability $\sin^{2} \eta$. In what follows, we will use the notation $c\equiv \cos \eta$ and $s \equiv \sin \eta$ for brevity.

If the state $\xi$ of the environment qubits corresponds to a Bloch vector $\vec{u}$, a single collision with the environment qubit induces the map $\vec{r}\rightarrow \vec{r}' = c^{2} \vec{r} + s^{2} \vec{u} - cs \, \vec{u}\times  \vec{r}$ on the Bloch vector of $S$. The corresponding atomic map in matrix form can be written as:
\begin{equation}
{\mathcal E}(\vec{u}) = \begin{pmatrix}
1 & 0 & 0 & 0  \\
 s^{2} u_x & c^{2} &  c s u_z & - c s u_y \\
 s^{2} u_y & - c s u_z & c^{2} &  c s u_x \\
 s^{2} u_z &  c s u_y & - c s u_x & c^{2} 
\end{pmatrix},
\end{equation}
where $u_x$, $u_y$ and $u_z$ are components of $\vec{u}$. In~\cite{ziman2005description} ${\mathcal E}_n$ is computed by re-writing the matrix to a more convenient operator basis which makes it easy to take its $n^{\rm th}$ power. However since we will be considering a sum of maps, we have to express all of them in the same basis. We start by writing the map in the form
\[ {\mathcal E}(\vec{u}) = \left( \begin{array}{cc}
	1 & \vec{0} \\
	s^2 \vec{u} & cA
\end{array} \right), \]
where 
\[ A = c {\mathbb I}_3 + s B, \qquad B = \left( \begin{array}{ccc}
	0 & u_z & - u_y \\ -u_z & 0 & u_x \\ u_y & - u_x & 0 
\end{array} \right). \]
Using this form for ${\mathcal E}(\vec{u})$ we obtain 
\begin{equation}
	\label{eq:n-map1}
	{\mathcal E}_n(\vec{u}) = \left( \begin{array}{cc}
	1 & \vec{0} \\
	{\mathbf u}_{n}  & c^nA^n
\end{array} \right),
\end{equation}
where ${\mathbf u}_{n} = {\mathbf u}_{n-1}+c^{n-1} A^{n-1} s^2 \vec{u}$, ${\mathbf u}_1 \equiv s^2 \vec{u}$. A long but straightforward calculation yields, 
\begin{eqnarray}
	\label{eq:Amat1}
	c^n A^n & = & \big[c^2(c^2 + u^2 s^2)\big]^{\frac{n}{2}} \big[\cos (n \Omega_u) {\mathbb I} + \sin (n \Omega_u) B \big]  \nonumber \\
	&&     + \, \Big(c^{2n} - \big[c^2(c^2 + u^2 s^2)\big]^{\frac{n}{2}} \cos(n \Omega_u)\Big) \frac{\vec{u} \vec{u}^T}{u^2}, \quad
\end{eqnarray}
where $u \equiv |\vec{u}|$ and $\vec{u} \vec{u}^T$ is the $3 \times 3$ outer product matrix of $\vec{u}$ with itself.  Here we have introduced the $u$-dependent frequency $\Omega_u \equiv \tan^{-1}(su/c)$. Using Eq.~(\ref{eq:Amat1}) and noting that $B\vec{u} = \vec{0}$ and $(\vec{u} \vec{u}^T)\vec{u} = u^2 \vec{u}$, we have $c^{n} A^{n} s^2 \vec{u} = s^2 c^{2n} \vec{u}$. This means that
\begin{eqnarray*}
	{\mathbf u}_n & = &  {\mathbf u}_{n-1} + s^2 c^{2(n-1)} \vec{u} = s^2 \vec{u} \big[ 1 + c^2 + \ldots c^{2(n-1)}\big] \\
	& = &  \vec{u}\big( 1-c^{2n} \big),
\end{eqnarray*} 
where, in the last step, the geometric progression appearing in the previous step has been summed. 

To transition to the continuous time version of the map, we use $n= t/\tau$ for a fixed (small) value of $\tau$ and define the following parameters:
\begin{equation}
	\omega_u \equiv  \frac{\Omega_u}{\tau}, \quad 
	\Gamma  \equiv  - \frac{2}{\tau} \ln c, \quad
	\gamma_u  \equiv  -\frac{1}{\tau} \ln (\cos \Omega_u).
	\label{eq:pardefs}
\end{equation}	
In terms of these parameters, we have $c^{2n} = e^{-\Gamma t}$, $n\Omega_u = \omega_u t$ and $\big[c^2(c^2 + u^2 s^2)\big]^{\frac{n}{2}} = e^{-\Gamma t} e^{\gamma_u t} $. So in the continuous time limit we have
\begin{eqnarray*}
	c^n A^n \rightarrow {\mathcal A}_t & = & e^{-\Gamma t} \bigg( e^{\gamma_u t} \bigg[ \cos (\omega_u t) {\mathbb I} + \sin (\omega_u t) \frac{B}{u} \bigg] \\
	& & \qquad \qquad +\; \Big[ 1 - e^{\gamma_u t} \cos(\omega_u t)  \Big] \frac{\vec{u} \vec{u}^T}{u^2} \bigg),
\end{eqnarray*}  
and
\[ {\mathbf u}_n \rightarrow {\mathbf u}_t = \vec{u} \big(1-e^{-\Gamma t} \big). \]
So we have
\begin{equation}
	\label{eq:ContinousMap1}
	{\mathcal E}_t(\vec{u}) = \left( \begin{array}{cc}
	1 & \vec{0} \\
	\vec{u} \big(1-e^{-\Gamma t} \big)  & {\mathcal A}_t
	\end{array} \right).
\end{equation}

It is worth mentioning here in passing that ${\mathcal E}_t^{-1}(\vec{u})$ can be computed easily if ${\mathcal A}_t^{-1}$ can be found and in turn, ${\mathcal A}_t$ is of the form $C + \vec{v}\vec{v}^T$ which can be inverted using the Sherman-Morrion formula~\cite{sherman49a} as 
\[ {\mathcal A}_t^{-1} = C^{-1} - \frac{C^{-1} \vec{v}\vec{v}^T C^{-1}}{1+ \vec{v}^T C^{-1} \vec{v}}, \]
provided $C \propto a {\mathbb I} + b B$ is invertible and $1+\vec{v}^T C^{-1} \vec{v} \neq 0$. Using ${\mathcal E}_t^{-1}(\vec{u})$  and $d{\mathcal E}_t/dt$ one can find the generator of the dynamical map using Eq.~(\ref{eq:generator}). 

Our interest is however in maps of the form $\widetilde{\mathcal E}_t$ as in Eq.~(\ref{eq:SumMap1}). Individual elements of this map other than the first row which is $ [1, \, 0, \, 0,\, 0]$ are listed below with the identification $u_x = u_2$, $u_y = u_3$ and $u_z = u_4$
\begin{eqnarray*}
	\big[\widetilde{\mathcal E}_t \big]_{j1} & = & (1-e^{-\Gamma t})\langle u_j \rangle, \quad j,k,l=2,3,4 \\
	\big[\widetilde{\mathcal E}_t \big]_{jj} & = &  e^{-\Gamma t} \bigg[ \big\langle e^{\gamma_u t} \cos \omega_u t \big\rangle + \bigg\langle \frac{u_j^2}{u^2} \bigg\rangle \\
	&& \qquad \qquad \qquad - \bigg\langle e^{\gamma_u t} \cos (\omega_u t) \frac{u_j^2}{u^2} \bigg\rangle  \bigg] \\
	\big[\widetilde{\mathcal E}_t \big]_{jk} & = &  e^{-\Gamma t} \bigg[ \epsilon_{jkl} \bigg\langle \frac{u_l}{u}  e^{\gamma_u t} \sin \omega_u t \bigg\rangle + \bigg\langle \frac{u_j u_k}{u^2} \bigg\rangle \\
	&& \qquad \qquad \qquad - \bigg\langle e^{\gamma_u t} \cos (\omega_u t) \frac{u_j u_k}{u^2} \bigg\rangle  \bigg]. 
\end{eqnarray*}
These elements are all written in terms of mean values $\langle u_j \rangle$, $\langle u_j u_k/u^2 \rangle$, $\langle e^{\gamma_u t} \cos (\omega_u t) \rangle$, $\langle u_j e^{\gamma_u t} \sin (\omega_u t)/u \rangle$ and $\langle e^{\gamma_u t} \cos (\omega_u t) u_j u_k/u^2 \rangle$ where 
\[ \langle x_u \rangle \equiv \int d^3 u \, p(\vec{u}) x_u. \]
In those cases where $\widetilde{\mathcal E}_t$ is invertible as well, we can find the generator of the evolution using Eq.~(\ref{eq:generator}) by computing the instantaneous time derivative of $\widetilde{\mathcal E}_t$.  The generator obtained from the derivative above is in general time dependent. From the generator, one can write a master equation in the GKSL form using the relations in Eq.~(\ref{eq:GKSL3}). By diagonalizing the $3\times3$ matrix with elements $c_{\alpha \beta}(t)$ we can write the qubit master equation in the canonical form
\begin{eqnarray}
	\label{eq:canonical}
	\frac{d\;}{dt}\rho_S(t) & = &  -i\big[H_t,\, \rho_S(t)\big]  +\frac{1}{2} \sum_{\alpha=1}^3 \lambda_{\alpha}(t) \Big( \big[ \zeta_\alpha^t, \, \rho_S(t) \zeta_\alpha^t \big] \nonumber \\
	&& \qquad \qquad \qquad \qquad + [\zeta_\alpha^t \rho_S(t), \, \zeta_\alpha^t \big] \Big),
\end{eqnarray}
where $\lambda_\alpha(t)$ are three instantaneous decay rates and $\zeta_\alpha^t$ are time dependent traceless, mutually orthogonal operators that, in turn, can be written in terms of the Pauli matrices. 

Using the map $\widetilde{\mathcal E}_t$ and the master equation (\ref{eq:canonical}), we can explore the following aspects of the dynamics corresponding to various choices of $p(\vec{u})$:
\begin{enumerate}
	\item Is the dynamics Markovian or non-Markovian?
	\item Is the dynamics invertible at all times so that it is also divisible?
	\item If the dynamics is divisible, then is it CP-divisible or P-divisible? 
\end{enumerate}
Verifying that the maps are CP for all values of $t$ provides a check on the accuracy of our numerical computations. 

There are quite a few approaches for detecting and quantifying non Markovianity in open quantum evolution. We choose to take information back-flow approach~\cite{laine2010measure, RevModPhys.88.021002} wherein the increase in the value of a suitable distance measure between any pair of states of the system during the evolution is taken as a signature of non-Markovian evolution. A quantifier of non-Markovianity can also be constructed using this approach but in our case, detecting non-Markovian evolution is sufficient. For the distance measure we choose the trace distance between two states $\rho_1$ and $\rho_2$ of the system defined as 
\[ D(\rho_1, \rho_2) = \frac{1}{2} {\rm Tr}\Big[ \sqrt{(\rho_1 - \rho_2)^\dagger (\rho_1 - \rho_2)} \Big]. \] 

Invertibility of the map can be directly ascertained and computing the determinant of the mapping matrix yields a test of invertibility. If the map is invertible, then CP and P divisibility are tested using the instantaneous rates $\lambda_\alpha (t)$ appearing in Eq.~(\ref{eq:canonical}). If all $\lambda_\alpha(t)$ are positive for all times, then the dynamics is CP divisible~\cite{rivas2014quantum} while P divisibility demands the weaker condition that the sum of any two of the rates remains positive at all times~\cite{megier2017eternal, wissmann2015generalized}.

\begin{widetext}
\subsection{Numerical investigations}

The parameter $\tau$ sets the typical time scale of the collisions in our numerical investigations. Taking $\tau$ to be small means that the collisions happen rapidly. Since $\widetilde{\mathcal E}_t$ is a continuous time interpolation of the sequence of discrete collisions, it stands to reason that having more collisions between $0$ and $t$ provides more points to fit the interpolation to, thereby improving the quality of the same. However, since $\tau$ appears in the denominator in Eq.~(\ref{eq:pardefs}), making $\tau$ small is not a good choice for the stability of the numerical integrations we do below. Fortunately we can compensate for making $\tau$ larger by reducing net impact that each collision has on the system so that we are still interpolating a slowly varying quantity. This can be achieved by choosing $\eta$ to be small. In all our computations below, we therefore choose $\tau=1$ for convenience while at the same time choosing $\eta = 0.01$. 

\subsubsection{Gaussian probability distribution centered at the origin}

We choose the probability distribution over $\vec{u}$ as a three dimensional Gaussian of the form
\begin{equation}
	\label{eq:gaussian}
	p(\vec{u}) = {\mathcal N} \frac{1}{(2\pi)^{3/2} \delta_x \delta_y \delta_z} e^{-\frac{(u_x-x_0)^2}{2 \delta_x^2} }e^{-\frac{(u_y-y_0)^2}{2 \delta_y^2} }e^{-\frac{(u_z-z_0)^2}{2 \delta_z^2} },
\end{equation}
where ${\mathcal N}$ is a normalization constant appearing because the Gaussian is defined only within the unit ball. The Gaussian is centered at $(x_0,y_0,z_0)$ and in each of the three directions it has standard deviations $\delta_x$, $\delta_y$ and $\delta_z$ respectively. For our first case we choose the Gaussian to be centered at the origin, $x_0=y_0=z_0=0$. We  also choose the widths of the distributions to be identical in all three directions $(\delta_x=\delta_y=\delta_z = \delta)$. First we investigate whether the evolution is Markovian on not. In Fig.~\ref{fig1a}-A, the trace distance $D(\rho_1, \rho_2)$ between states $\rho_1 = ({\mathbb I}+\sigma_z)/2$ and $\rho_2 = {\mathbb I}/2$ is plotted as a function of time for three choices of the width $\delta$ of the Gaussian. We see that for wide Gaussians corresponding to $\delta=0.3$ and $\delta=0.1$ the evolution described by Eq.~(\ref{eq:canonical}) is non-Markovian. The spatial grid used for the numerical integrations in the $u_x$, $u_y$, $u_z$ space is of size $0.05$. So by choosing $\delta = 0.01$ in the third case, we have a very good approximation to the case where $p(\vec{u})$ is a delta function corresponding to homogenization to the fully mixed state. In this ideal case we have a dynamical map arising from a single collision model which is Markovian. We see that when $\delta=0.01$, the separation between the two states considered is monotonically decreasing indicating that the evolution is indeed Markovian in this case as expected. 
\begin{figure}[!htb]
	\resizebox{13 cm}{4.5cm}{\includegraphics{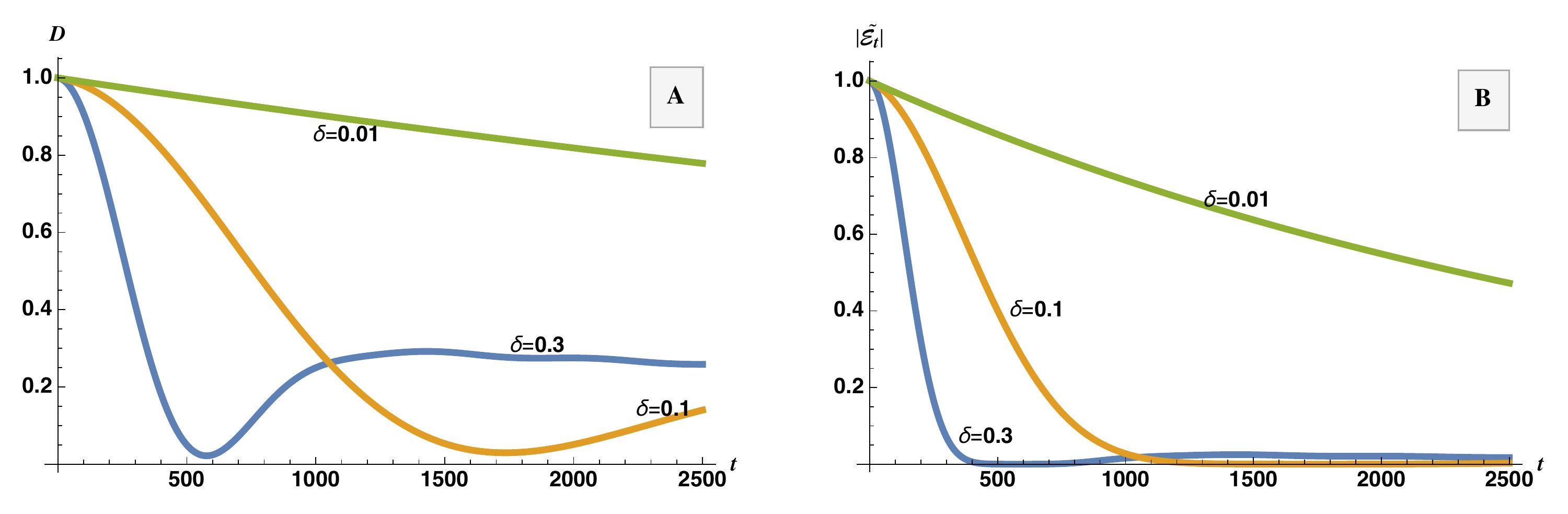}}
	\caption{\label{fig1a} The figure on the left (A) shows the separation of two states as a function of time for three choices of the width of the Gaussian. The figure on the right (B) shows the determinant of $\widetilde{\mathcal E}_t$ as a function of time.}
\end{figure}

Fig.~\ref{fig1a}-B shows the determinant of $\widetilde{\mathcal E}_t$ as a function of time. The determinant remains positive for all times indicating that the map is invertible in this case at all times. Even though the determinant becomes very small as seen in the graphs, it never touches zero except asymptotically. Invertibility was also verified independently by directly computing the inverse of the map. To check the accuracy of our numerical investigations, we compute the ``$B$-matrix" form of the dynamical map~\cite{PhysRev.121.920, Nielsen00a} and compute the eigenvalues of $B$ as well as its trace. For all qubit maps we have ${\rm Tr}(B)=2$ and if all the eigenvalues of the $B$-matrix are positive then the map is CP. The four eigenvalues $b_j$ of $B$ as well as ${\rm Tr}(B)$ is plotted in Fig.~\ref{fig1b} and we find that indeed the trace of $B$ is always $2$ and all its eigenvalues are positive at all times for all three choices of $\delta$ as expected. This serves to cross check the accuracy and stability of our numerical integrations.  
\begin{figure}[!htb]
	\resizebox{17cm}{4.5cm}{\includegraphics{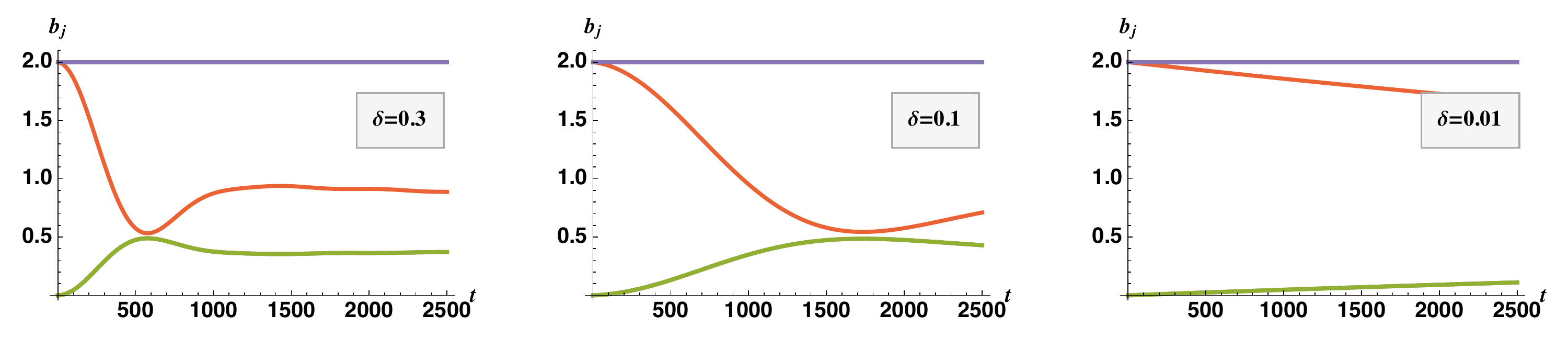}}
	\caption{\label{fig1b} The four eigenvalues of the mapping matrix when $\widetilde{\mathcal E}_t$ is written in the ``$B$-matrix" form are plotted for the three cases $\delta = 0.3$, $\delta = 0.1$ and $\delta=0.01$. Also shown are the traces (purple horizontal lines on top of each plot) of the corresponding numerically computed $B$-matrices that has to have the value of $2$ for all qubit maps. All the eigenvalues $b_j$ of the $B$-matrix being positive indicates that the map $\widetilde{\mathcal E}_t$ is CP in all three cases and for all values of $t$. In all three cases there are only two distinct eigenvalues each of which is doubly degenerate.}
\end{figure}

To see if the invertible, completely positive dynamics is CP-divisible or not, we find the rates in Eq.~(\ref{eq:canonical}) for the three choices of $\delta$. We find that the isotropy of $p(\vec{u})$ is reflected in the rates as well with the three rates for each choice of $\delta$ being the same. The rates corresponding to the three cases are plotted in Fig.~\ref{fig1c}. When the dynamics is non-Markovian corresponding to $\delta = 0.3$ and $\delta = 0.1$, the rates show a strong time dependence while in the approximately Markovian case with $\delta = 0.01$ the rates are almost constant. We further see that the dynamics is not CP-divisible when $\delta = 0.3$ or $\delta = 0.1$ since the rates are not always positive. The sum of any two of the rates does not remain positive either for the wide distributions indicating lack of P-divisibility.  
\begin{figure}[!htb]
	\resizebox{8.5cm}{5.0cm}{\includegraphics{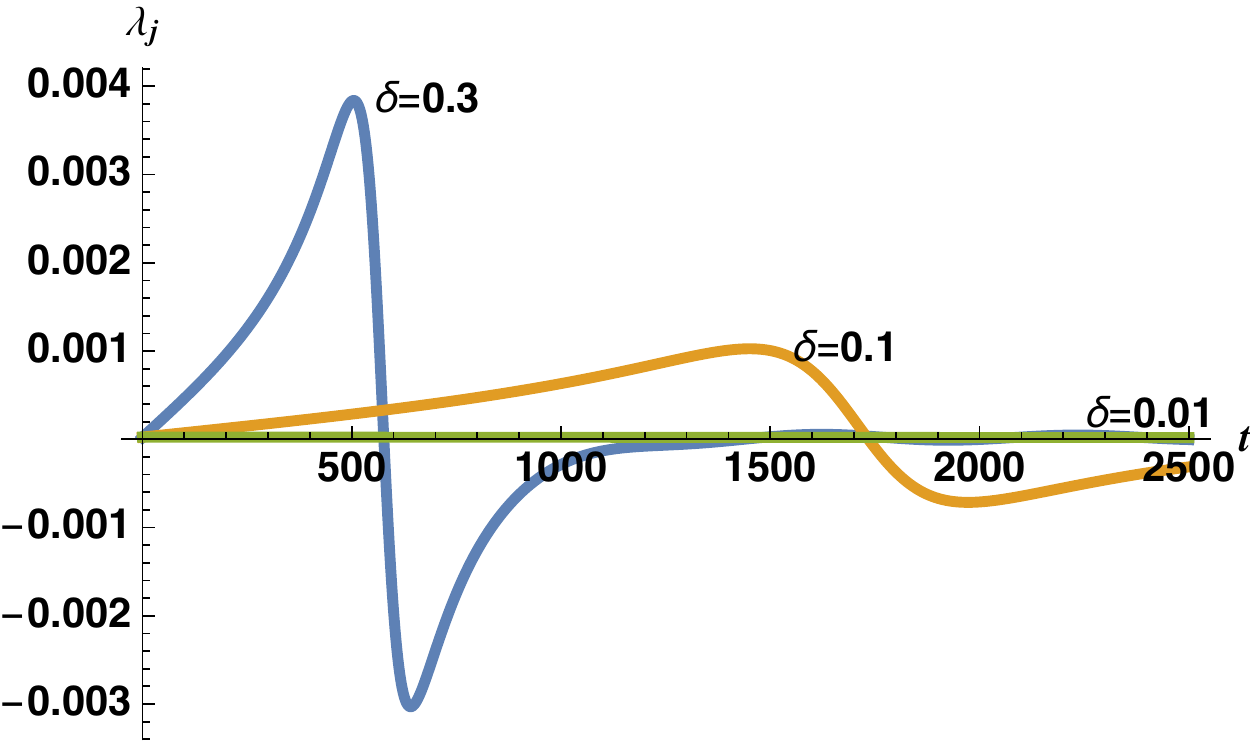}}
	\caption{\label{fig1c} The rates $\lambda(t)$ from Eq.~(\ref{eq:canonical}) plotted as a function of time for three choices of $\delta$. For each $\delta$ all three rates are the same and so only one each is plotted.} 
\end{figure}

Now we consider an anisotropic distribution $p(\vec{u})$ again centered at the origin but having significant width along the $u_z$ direction only. This is motivated by the observation that when $\vec{u} = (0,0,u_z)$, the dynamical map in Eq.~(\ref{eq:ContinousMap1}) has the form~\cite{ziman2005description}
\[ {\mathcal E}_t(u_z) = \left( \begin{array}{cccc}
	1 & 0 & 0 & 0 \\
	0 & e^{-\Gamma_u't} \cos (\omega_u t) & e^{-\Gamma_u't} \frac{u_z}{u} \sin (\omega_u t) & 0 \\
	0 & - e^{-\Gamma_u't} \frac{u_z}{u}\sin (\omega_u t)  & e^{-\Gamma_u't} \cos (\omega_u t)  & 0 \\
	u_z(1-e^{-\Gamma t}) & 0 & 0 & e^{-\Gamma t} 
\end{array} \right), \]
where $\Gamma_u' \equiv \Gamma - \gamma_u$. From the form above, it is easy to see that two of the eigenvalues of the map $[{\mathcal E}_t(u_z) + {\mathcal E}_t(-u_z)]/2$ are zero whenever $\cos(\omega_u t) = 0$ making it non-invertible for the corresponding values of $t$. Here we consider a more general convex combination of the maps ${\mathcal E}_t(u_z)$ by choosing $\delta_x = \delta_y = 0.01$ and $\delta_z=0.7$. 

\begin{figure}[!htb]
	\resizebox{13 cm}{4.5cm}{\includegraphics{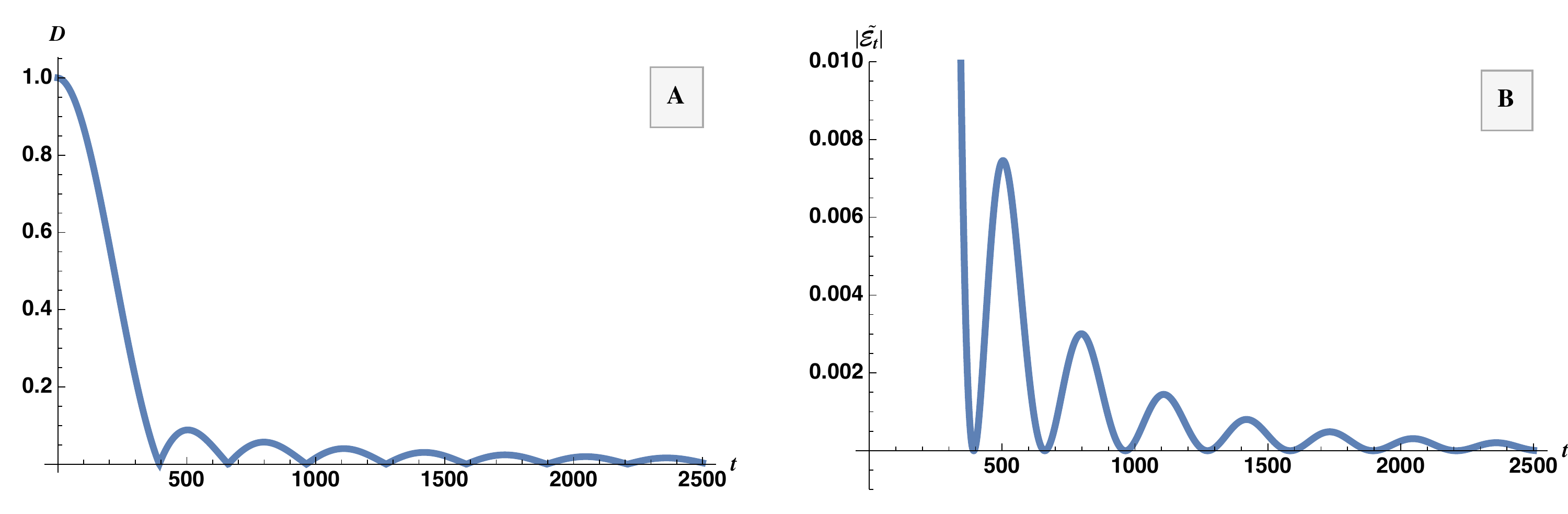}}
	\caption{\label{fig2a} The figure on the left (A) shows the separation of two states as a function of time when $\delta_x = \delta_y = 0.01$ and $\delta_z=0.7$. The figure on the right (B) shows the determinant of $\widetilde{\mathcal E}_t$ as a function of time for this case.}
\end{figure}

In Fig.~\ref{fig2a}-A, the trace distance between the states $({\mathbb I} + \sigma_x)/2$ and ${\mathbb I}/2$ is plotted as a function of time. We see that the separation behaves non-monotonically, indicating non-Markovian evolution. The discontinuity in the slope of graph of the separation is because the Bloch sphere of states gets reflected through the $z$-axis periodically during the course of the evolution. Fig~\ref{fig2a}-B shows the determinant of $\widetilde{\mathcal E}_t$ and we see that the determinant vanishes periodically making the map non-invertible at those points in time. As before the eigenvalues and trace of the $B$-matrix serve as a check on the accuracy of the numerics and we find that the map indeed is CP at all times but the plots of its eigenvalues are avoided for brevity. 

\begin{figure}[!htb]
	\resizebox{13 cm}{4.5cm}{\includegraphics{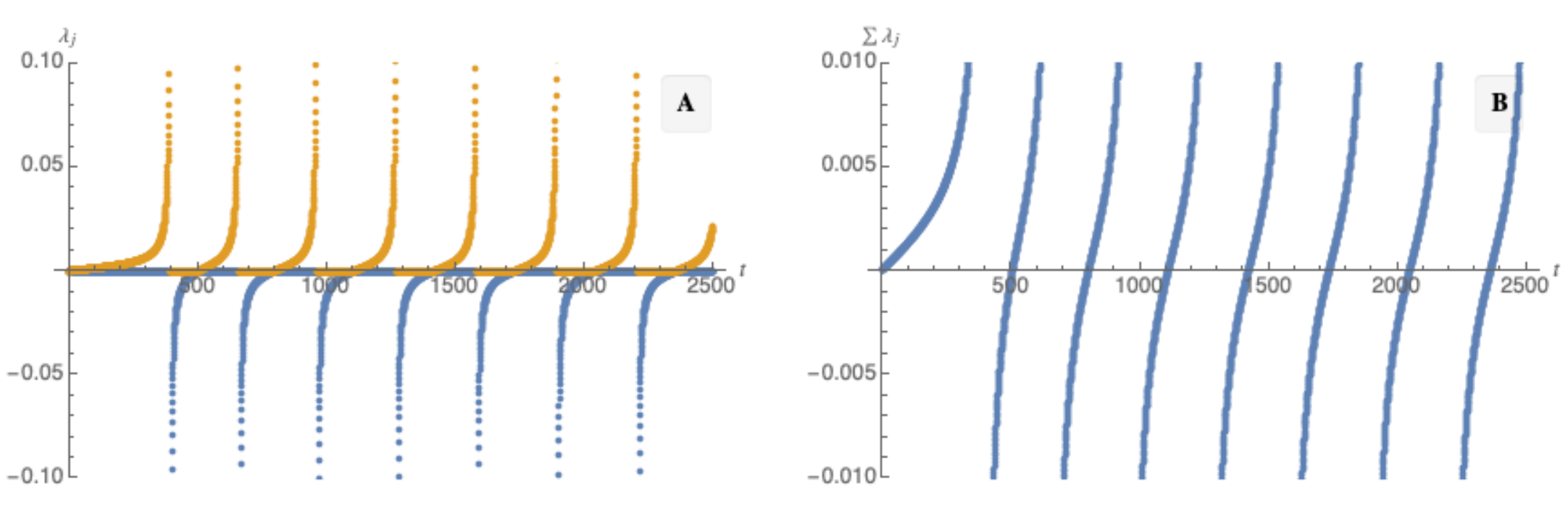}}
	\caption{\label{fig2c} The figure on the left (A) shows the two non-zero rates from Eq.~(\ref{eq:canonical}) as a function of time. The figure on the right (B) shows the sum of the two rates. Both figures are for a Gaussian distribution centered at the origin with $\delta_x = \delta_y = 0.01$ and $\delta_z=0.7$}
\end{figure}

We can explore the divisibility properties of the map by computing the rates in Eq~(\ref{eq:canonical}) for all values of $t$ for which the dynamical map is invertible. Fig.~\ref{fig2c}-A shows the two non-zero rates that are obtained. The third rate corresponding to contraction or expansion along the long axis of the distribution is effectively zero. The rates diverge as expected at those points where the inverse does not exist and one of the rates is always negative elsewhere indicating that the dynamics is not CP-divisible. The map is not P-divisible either since one of the rates is zero and one is always negative and so the sum of the two is always negative. Fig.~\ref{fig2c}-B shows the sum of the two non-zero rates which also goes negative periodically.

\subsubsection{Gaussian probability distribution centered not at the origin}

We now consider an isotropic Gaussian distribution with $\delta = 0.3$ centered at the point $(0.3,0,0)$ as $p(\vec{u})$. Fig~\ref{fig3a} summarizes the results in this case. The separation between $({\mathbb I} + \sigma_z)/2$ and ${\mathbb I}/2$ shows non-monotonic behavior in Fig~\ref{fig3a}-A indicating non-Markovian evolution. The determinant is always positive and the dynamical map is verified to be invertible at all times. The three rates from the master equation is plotted in Fig~\ref{fig3a}-C and the pair-wise sums of the three rates in Fig~\ref{fig3a}-D. We find that the evolution is neither CP nor P-divisible for this choice of $p(\vec{u})$ since both the individual rates as well as their pair-wise sums are not positive semi-definite for all $t$. 

\begin{figure}[!htb]
	\resizebox{13cm}{9cm}{\includegraphics{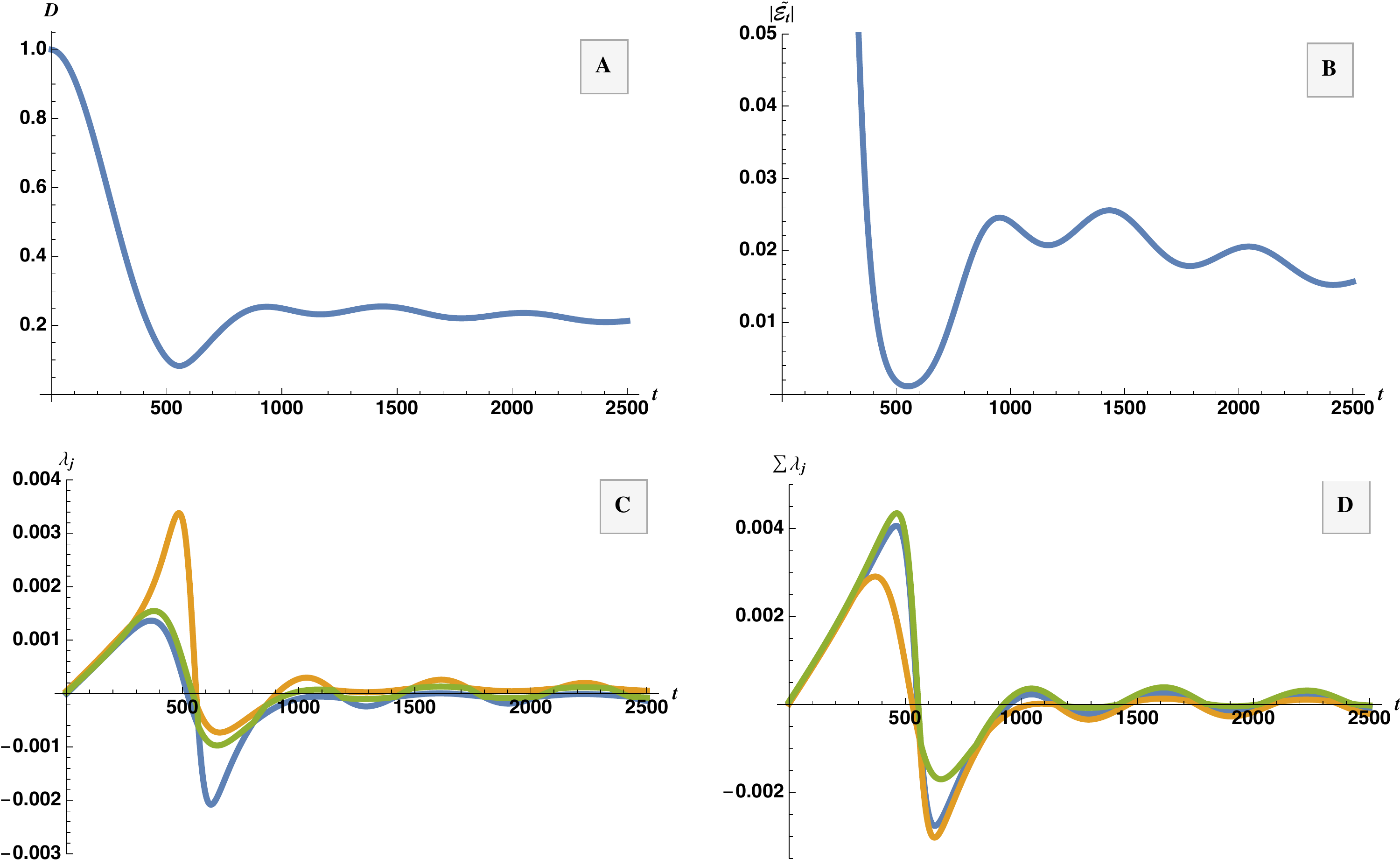}}
	\caption{\label{fig3a} Plots of $p(\vec{u})$ centered at $(0.3,0,0)$ with $\delta_x=\delta_y=\delta_z=0.3$. Panel (A) shows the trace distance between two states as a function of time and panel (B) shows the determinant of $\widetilde{\mathcal E}_t$. The three rates appearing in the master equation are shown in panel (C) while the pair-wise sums of the three rates is in (D).}
\end{figure}

For completeness we consider an anisotropic distribution as well centered at $(0.3,0,0)$. We consider a disk shaped distribution with $\delta_x=0.01$, $\delta_y=\delta_z=0.3$. The results are summarized in Fig.~\ref{fig4a}. From the evolution of the trace distance between $({\mathbb I} + \sigma_z)/2$ and ${\mathbb I}/2$ we again see that the evolution is non-Markovian. We also find that the evolution is invertible but it is neither CP nor P divisible in this case as well. 

\begin{figure}[!htb]
	\resizebox{13cm}{9cm}{\includegraphics{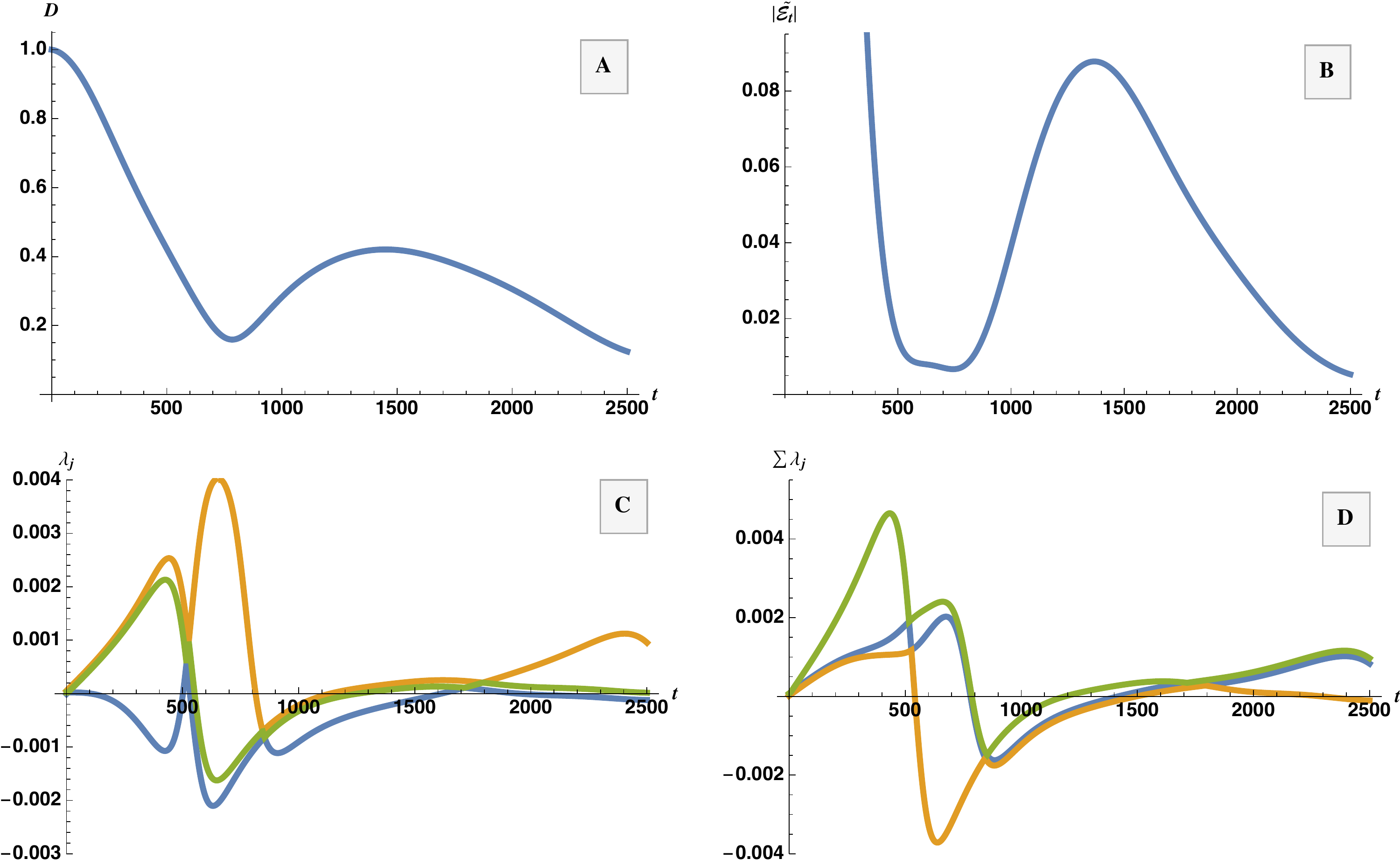}}
	\caption{\label{fig4a} Plots of $p(\vec{u})$ centered at $(0.3,0,0)$ with $\delta_x=0.01$, $\delta_y=\delta_z=0.3$ Panel (A) shows the trace distance between two states as a function of time and panel (B) shows the determinant of $\widetilde{\mathcal E}_t$. The three rates appearing in the master equation are shown in panel (C) while the pair-wise sums of the three rates is in (D).}
\end{figure}

\section{Conclusions \label{conclusions}}

We have seen that considering dynamical maps that are convex combinations of the CP-divisible, invertible and Markovian maps that arise from collision models can lead to a wide variety of maps which are non-Markovian, non-invertible and not CP-divisible either. These dynamical maps and corresponding master equations describing it are guaranteed to be completely positive in this case and there is no need to put in additional constraints as is often the case when phenomenological master equations are constructed. We analyzed the particular case of the qubit homogenization map in detail and observed that convex combinations corresponding to wide distributions typically lead to non-Markovian dynamics. We saw cases where the dynamics is invertible and non-invertible as well as cases wherein the dynamics is CP-divisible, P-divisible or neither. Obtaining a closed form expression for ${\mathcal E}_n(\vec{u})$ was an important step that allowed us to construct the arbitrary combinations of such maps. Different choices of $p(\vec{u})$ revealed a rich structure with several different features for the corresponding dynamics. An exhaustive enumeration of the properties of the dynamical maps corresponding to different choices of $p(\vec{u})$ remains to be done. Also, the primitive map we have studied is restricted to the qubit homogenization map using the swap operator in Eq.~(\ref{eq:swap}). The same analysis can be extended to other unitaries coupling the system to the environment particles. Considering maps on more general systems than qubits is another avenue for generalization.

Viewing non-Markovian evolution as arising from a convex combination of collision model based CP-divisible dynamical maps provides a kind of unravelling of the dynamics that is intuitively understandable. It furnishes a picture of environment particles in various states interacting with the system with relative frequencies fixed by the distribution $p(\vec{u})$. The original understanding of Markovian open quantum dynamics developed by Gorini, Kossakowski and Sudarshan can therefor have a bearing on the study of a non-Markovian dynamics as well. However such a picture would be particularly useful if a systematic method for deconstructing general families of dynamical maps corresponding to non-Markovian evolution into such convex combinations of collision model based dynamical maps is available. This question can be addressed through a detailed exploration of the dynamics arising from various choices of $p(\vec{u})$ and different types of interactions between the system and the environment particles. Such detailed explorations is beyond the scope of the present work and it will be left for the future. 

\section*{Acknowledgements}
Anil Shaji acknowledges the support of the Department of Science and Technology, Government of India through grant no. EMR/2016/007221. The authors thank Dr. Jyrki Piilo for useful discussions during exchange visits facilitated by an Erasmus Plus Mobility grant from the EU. 
\end{widetext}

\bibliography{masterequation}

\end{document}